\documentclass[sigconf]{acmart}

\usepackage{amstext}        
\usepackage{mathtools}
\usepackage{subcaption}
\usepackage{algorithm}
\usepackage{algorithmic}
\usepackage{multirow}
\usepackage{nicefrac}
\usepackage{textcomp}

\newcommand{\Var}{\mathrm{Var}}
\newcommand{\TypeSet}{\Theta}
\newcommand{\Types}{\boldsymbol{\theta}}
\newcommand{\mathds}[1]{\mathbb{#1}}
\newtheorem{assumption}{Assumption}
\newtheorem{remark}{Remark}

\copyrightyear{2026}
\acmYear{2026}
\setcopyright{cc}
\setcctype{by}
\acmConference[ICAIF '26]{7th ACM International Conference on AI in Finance}{November 14--17, 2026}{Milan, AA, Italy}
\acmBooktitle{7th ACM International Conference on AI in Finance (ICAIF '26), November 14--17, 2026, Milan, AA, Italy}
\acmDOI{10.1145/3858063.3858711}
\acmISBN{979-8-4007-3087-0/2026/11}

\begin{document}

\title{Fair Contracts in Principal-Agent Games with Heterogeneous Types}

\author{Jakub Tłuczek}
\email{jakub.tluczek@unine.ch}
\orcid{0009-0006-9970-2521}
\affiliation{%
  \institution{Université de Neuchâtel}
  \city{Neuchâtel}
  \country{Switzerland}
}

\author{Victor Villin}
\email{victor.villin@unine.ch}
\orcid{0009-0004-6778-1494}
\affiliation{%
  \institution{Université de Neuchâtel}
  \city{Neuchâtel}
  \country{Switzerland}
}

\author{Christos Dimitrakakis}
\email{christos.dimitrakakis@unine.ch}
\orcid{0000-0002-5367-5189}
\affiliation{
  \institution{Université de Neuchâtel}
  \city{Neuchâtel}
  \country{Switzerland}
}

\renewcommand{\shortauthors}{Tłuczek et al.}

\begin{abstract}
    Fairness is desirable yet challenging to achieve within multi-agent systems, especially when agents differ in latent traits that affect their abilities. This hidden heterogeneity often leads to unequal distributions of wealth, even when agents operate under the same rules. Motivated by real-world examples, we propose a framework based on repeated principal-agent games, where a principal learns to offer adaptive contracts to agents. By leveraging a simple yet powerful contract structure, we show that a fairness-aware principal can learn homogeneous linear contracts that equalize outcomes across agents in a sequential social dilemma. Importantly, this fairness does not come at the cost of efficiency: our results demonstrate that it is possible to promote equity and stability in the system while preserving overall performance.
\end{abstract}



\maketitle

\section{Introduction}
\label{sec:intro}
Modern economies, at both macro and micro levels, constitute complex multi-agent systems. Interactions between agents often involve contracts designed to align incentives between parties, fitting the principal-agent model: one party (the principal) offers a reward to another (the agent) in exchange for a specific outcome. This framework applies to a wide range of real-world scenarios, from employment agreements to government subsidies.

While usually parties in the principal-agent model are assumed to be greedy with respect to their own wealth, this assumption may not hold in real-life settings. It has been shown~\cite{fehr_fairness_2007} that fairness concerns about the other party can be a potential decisive factor in contracting schemes. The boss might be inclined to offer bigger contracts to ensure that the agent stays motivated, while the agent might exert more effort anticipating bigger contracts rewarding him in the future.

Contract design in principal-agent settings poses a significant challenge. Self-interested principals tend to disregard the agents' wealth. From the agent's perspective, this can mean accepting unfavorable conditions or rejecting them altogether, resulting in either exploitation or stagnation. If contracts are not carefully constructed, inequality can also appear between agents. Encouraging principals to account for the global health of the system should enhance social interactions and lead to more prosperous outcomes.
 
These insights extend naturally to the domain of Multi-Agent Reinforcement Learning (MARL), where principals and agents are independently seeking to maximize their rewards through strategic decision making. Within this setting, a principal can serve as a central planner (who still potentially participates in the game), offering contracts that agents may choose to accept or reject. By shaping reward structures through contract design, the principal can steer agents toward more cooperative and equitable behavior, ultimately promoting system-wide efficiency and fairness.

\subsection{Main Contributions}
\label{subsec:contributions}
Our aim is to design adaptive contracts to obtain equitable outcomes between agents, while maintaining a high utility for the principal. In particular:
\begin{enumerate}
  \item In Section~\ref{sec:prelims}, we formalize the principal-agent model through MARL, and state the objectives for the principal and agents.
  \item In Section~\ref{sec:learning_contracts}, we interpret our setup through the lens of contract theory (CT), establishing theoretical grounding for contract validity. We then study learning simple linear contracts, using policy gradient methods. 
  \item We propose regularizing the principal's objective for contract design with a specific fairness function in Section~\ref{sec:regularization}. We explore two such functions: the first one adapts prior work~\cite{ivanov_principal-agent_2024} and incorporates the wealth of agents into the principal's objective, while the second explicitly penalizes inequality, measured by variance.
  \item We verify our methods on two sequential social dilemmas and explore the emerging behaviors adopted by agents and principals (Section~\ref{sec:experiments}). Our results demonstrate that variance regularization can lead to learning contracting strategies that yield a stable and fair principal-agent system.
\end{enumerate}

\section{Related Work}
\label{sec:relwork}
\textbf{Multi Agent Reinforcement Learning} studies multiple agents acting in a shared environment, either centrally orchestrated or learning independently~\cite{huh_multi-agent_2024,zhang_multi-agent_2021}. Optimality is problem-dependent, ranging from equilibria (Nash, SPE, Stackelberg), Pareto optimality, welfare to fairness~\cite{albrecht_multi-agent_2024}. We optimize for fairness among the independently acting agents whose behavior is mediated by a contract mechanism.

\textbf{Incentive design} is realized by providing external rewards to align agents with a particular goal. It is one of the fundamental challenges in the field of mechanism design~\cite{nisan_algorithmic_2007}. In scenarios with multiple agents, we may modify the rewards to incentivize agents to take desired actions~\cite{paccagnan_utility_2022} or to induce cooperative behavior in the population~\cite{danassis_ai-driven_2023,yang_learning_2020,zheng_ai_2020}. In our framework we are using homogeneous contracts, offered to all agents, to induce fairness in the system.
 
\textbf{Sequential social dilemmas} (SSD) are Markov games where agents can exhibit various degrees of cooperation~\cite{leibo_multi-agent_2017}. In their simplest form, they are repeated normal form games such as the iterated Prisoner's Dilemma. We test our solution on two SSD: Coin Game~\cite{foerster_learning_2018} and Gather-Trade-Build~\cite{zheng_ai_2020}, where agents do not have an underlying reason to cooperate. Using simple, linear contracts to induce cooperation in a heterogeneous population is one of our main contributions.

\textbf{Contract theory} is a well established subdomain in the field of economics~\cite{bolton_contract_2005,holmstrom_aggregation_1987,laffont_theory_2009,salanie_economics_2005}, which has recently started to attract attention in computer science~\cite{alon_bayesian_2023,babaioff_combinatorial_2006,duetting_algorithmic_2024,duetting_simple_2019,duetting_complexity_2020,guruganesh_contracts_2021}. The principal-agent model with multiple agents has already been a topic of several groundbreaking works~\cite{castiglioni_multi-agent_2023,duetting_multi-agent_2022,duetting_multi-agent_2024}. Informational assumptions vary and include bandit games~\cite{scheid_incentivized_2024}, learning in repeated settings~\cite{guruganesh_contracting_2024,dogan_repeated_2023} and hidden rewards settings~\cite{dogan_estimating_2023}. 

Our setting builds on the framework introduced in~\cite{ivanov_principal-agent_2024}, but we have a different focus. Firstly, our objective is to ensure fairness between the agents while maintaining social welfare, rather than finding an equilibrium. Secondly, all policies are learned simultaneously from scratch. Finally, agents in our settings are heterogeneous, each having a private type that affects the rewards. 

\textbf{Typed contracts} form a subset of contract theory, where contracting agents have hidden types. Two common approaches are \textit{menus of contracts}~\cite{castiglioni_designing_2022}, in which agents select both a type and an action,  and \textit{type-soliciting contracts}~\cite{alon_contracts_2021}, in which agent declares his types and the principal responds with an offer~\cite{duetting_algorithmic_2024}. Our framework falls in neither: our agents neither select nor disclose their type, with the challenge being offering the same contract to all of them.

\textbf{Fairness in multi-agent systems} is a notion that can be conceptualized in multiple ways. We focus on equity, ensuring that each agent obtains comparable benefits from the system. Related notions include equity-driven learning when agents observe each other's rewards~\cite{jiang_learning_2019}, cake-cutting allocations based on private valuations~\cite{brandt_handbook_2016}, proportional fairness~\cite{ju_achieving_2023}, and individual fairness via a Lipschitz condition~\cite{dwork_fairness_2011}. The latter two can only be evaluated under perfect information, which makes them not suitable for our setting due to the fact that agents' types are hidden, making equity a more practical objective.

Recently, the topic of envy-freeness in contracts has been explored by~\cite{castiglioni_fair_2025}. It adds a task allocation: the principal offers each agent a task with a corresponding linear contract. The system is deemed fair if no agent prefers an allocation of another. In our setting, agents implicitly share a single task and the contract is homogeneous to all of them.

\section{Preliminaries}
\label{sec:prelims}

\paragraph{Markov Games.}  A finite horizon $n$-player Markov game can be formalized as a tuple $\mathcal{M}=(\mathcal{S}, \mathcal{A}, \mathcal{P}, R, T, s_0)$, where $\mathcal{S}$ is a set of states, $\mathcal{A} = A^1 \times ... \times A^n$ is a set of discrete actions for each player, $\mathcal{P}: \mathcal{S} \times \mathcal{A} \rightarrow \Delta^{\mathcal{S}}$ is a transition function, $R: \mathcal{S} \times \mathcal{A} \times \mathcal{S} \times \{1, \dots, n \} \rightarrow \mathbb{R}^+$ is a reward function, $T$ is a horizon, and $s_0$ an initial state.

At each time step $t$, all players observe the current state $s_t\in \mathcal{S}$, select actions $\mathbf{a}_t = (a^1_t, \dots, a^n_t) \in \mathcal{A}$, and receive individual rewards $\mathbf{r}_t = (r^1_t, \dots, r^n_t)$. The following time step, the game transitions to a new state $s_{t+1} \sim \mathcal{P}(s_t, \mathbf{a}_t)$. This process repeats until the horizon $T$ is reached. Each player $i$ follows a policy $\pi^i$ that maps their observations to a distribution over actions in $A^i$.

\paragraph{Heterogeneous Principal-Agent Markov Games.}
To account for contracts in this framework, we define a heterogeneous principal-agent Markov game \cite{ivanov_principal-agent_2024} as an $(n+1)$-player extension of an underlying $n$-player game $\mathcal{M}$. Formally, we denote it as $\mathcal{M}_\text{pa}= (\mathcal{M}, B, \mathcal{A}_\text{a}, \Types, R_\text{p}, R_\text{a})$, where the first $n$ players are \emph{agents}, and the additional player indexed $p = n+1$ acts as the \emph{principal}. In our setting, the principal gives payments to agents that are a function of the reward they collect. Formally, $B$, the action space of the principal, is the set of contracts $b : \mathbb{R}^+ \rightarrow \mathbb{R}^+$, mapping from rewards to payments. The joint action space of agents is $\mathcal{A}_\text{a} = A^1_\text{a} \times \dots \times A^n_\text{a}$, where $A_\text{a}^i = A^i \cup \{\varnothing\}$ augments the agent's action space to include a game-ending "reject contract" action, denoted by $\varnothing$.

To capture heterogeneity among agents, each agent $i$ is endowed with a hidden type $\theta^i \in \TypeSet$, with $\Types$ the vector of all agent types.  Following~\cite{zheng_ai_2020}, the type is simply a scalar determining their efficiency.

Consequently, the contractual reward function for agents is defined as:
\begin{align}
  \label{eq:basic_reward_agent}
    R_\text{a}(s_t, \mathbf{a}_t, s_{t+1}, b_t, i) = \left( b_t(\theta^i r_t^i) - c \right) \cdot \mathds{I}[a^i_t \neq \varnothing], \quad i \neq p,
\end{align}
where $b_t : \mathbb{R}^+ \to \mathbb{R}^+$ is a contract function that specifies the payment based on the original reward, and $c$ is a fixed cost incurred for taking a non-null action. The dependency on states, actions, and next states comes from the underlying reward function. Agents who accept the contract receive a reward determined by $b_t$ and incur a cost $c$. Agents who reject the contract by playing the null action receive no reward.
In contrast, the principal's reward corresponds to the difference between the agents' raw contributions and their payments, i.e., the net surplus:
\begin{align}
  \label{eq:basic_reward_principal}
  R_\text{p}(s_t, \mathbf{a}_t, s_{t+1}, b_t) = \sum_{i=1}^{n} \left(\theta^i r_t^i - b_t(\theta^i r_t^i) \right) \cdot \mathds{I}[a^i_t \neq \varnothing].
\end{align}
 
The joint policy includes both the principal's policy and the agents' policies $\boldsymbol{\pi} = (\boldsymbol{\pi}_\text{a}, \pi_\text{p}) = (\pi^1_\text{a}, \dots, \pi^n_\text{a}, \pi_\text{p})$. The principal-agent interaction unfolds in three steps:
\begin{enumerate}
  \item The principal proposes a unique and homogeneous contract to all agents upon observing state $s$, $b \sim \pi_\text{p}(\cdot \mid s)$.
  \item The agents observe the common state $s$ and respond to the proposed contract $b$ by sampling an action $a^i \sim \pi^i_\text{a}(\cdot \mid s, b)$ from their policy.
  \item A new state $s'$ is generated from $\mathcal{M}_\text{pa}$, the agents pay their costs $c$, and obtain their contractual rewards $b(\theta^i r^i)$.
\end{enumerate}
This framework allows the principal to shape the system dynamics by offering incentive-aligned contracts, while agents respond strategically based on their preferences and the contract terms.

\begin{definition}
  \label{def:wealth}
  The $i$-th player's \emph{wealth} $w^i_t$ at time $t$ is defined as the total return until time $t$, starting from the fixed initial state $s_0$:
  \begin{align}
  \label{eq:wealth}
    w^i_t = \sum_{k=0}^{t} R_\text{a}^i(s_k, \mathbf{a}_k, s_{k+1}, b_k, i) \, .
  \end{align}
\end{definition}

\begin{definition}
  \label{def:value}
  In the principal-agent Markov game $\mathcal{M}_\text{pa}$, \emph{the value function} of an agent $i$ at time $t$ in state $s$ under contract $b$ is the expected cumulative reward obtained from that point onward, under the joint policy $\boldsymbol{\pi}$. In other words, it is the difference between expected final wealth and wealth at time $t$
  \begin{align}
  \label{eq:agents_value}
  V_t^i(s) &\coloneqq \mathbb{E}^{\boldsymbol{\pi}}_{\mathcal{M}_\text{pa}} \left[\sum_{k=t}^{T-1} R_\text{a}^i(s_k, \mathbf{a}_k, s_{k+1}, b_k, i) \;\middle|\; s_t = s \right] \notag\\
  & \;= \mathbb{E}^{\boldsymbol{\pi}}_{\mathcal{M}_\text{pa}} \left[ w_{T-1}^i  - w_{t-1}^i \right], \quad i \neq p \, .
\end{align}
\end{definition}

The value function and wealth can analogously be defined for the principal by replacing $R_\text{a}^i$ with $R_\text{p}$. 

\begin{assumption}
  \label{ass:maximizing}
  Each player maximizes their \emph{expected wealth}:
  \begin{align}
    \max_{\pi_a^i} \mathbb{E}^{\boldsymbol{\pi}}_{\mathcal{M}_\text{pa}} \left[ w_{T-1}^i \right].
  \end{align}
\end{assumption}

We denote the set of wealth over all players, the principal, and agents as $\mathcal{W}_t = \{w^1_t, \dots, w^n_t, w^p_t \}$, and we define the system's \emph{welfare} as $W_t \coloneqq \sum_{w_t \in  \mathcal{W}_t} w_t$.

\section{Learning Contracts}
\label{sec:learning_contracts}

To effectively learn the principal's policy for fair contract design, we must first formalize the core properties that define a valid CT setting, including the conditions under which agents should accept or reject contracts. In Section~\ref{subsec:contracts}, we analyze the CT literature to assess whether our framework satisfies the standard constraints. Then, in Section~\ref{subsec:fair_contracts}, we focus on a specific case where contracts are modeled as simple linear functions of the rewards perceived by agents. Finally, in Section~\ref{sec:policy_gradients} we present the learning algorithm for principal-agent policy gradients.

\subsection{Contract Theory}
\label{subsec:contracts}

The principal-agent model~\cite{bolton_contract_2005,duetting_simple_2019,holmstrom_aggregation_1987,salanie_economics_2005} is characterized by two parties aligning their interests with the help of contracts. The principal offers a contract which specifies payments conditioned on \emph{outcomes}. The agent then decides whether to accept the contract or not: if the contract is accepted, the agent takes a hidden action, bearing its cost. Although the principal receives the reward for the outcome, the agent is rewarded according to the contract agreed on beforehand. In the principal-agent Markov game defined in Section~\ref{sec:prelims}, we assume that the outcomes are identical to the rewards collected by the agents.

To prevent trivial or degenerate solutions, contract theory imposes three standard constraints~\cite{duetting_algorithmic_2024}:

\begin{definition}
  \label{def:ic}
  \emph{Incentive Compatibility} (IC): Agents are rational and aim to maximize their own wealth. A contract is \emph{incentive compatible} if the agent's best response, i.e., the action that maximizes wealth under the contract, is desired by the principal. Actions that satisfy this condition are called \emph{implementable}.
\end{definition}

\begin{definition}
  \textit{Limited liability} (LL): Contractual payments must always flow from the principal to the agent. Agents must never have to pay the principal. 
  \begin{align}
    \forall_{i,t} \quad b_t \left( \theta^i r_t^i \right) \geq 0.
  \end{align}
\end{definition}

\begin{definition}
  \emph{Individual Rationality} (IR): Agents should reject contracts that yield negative expected returns. Formally, agent $i$ accepts a contract $b_t$ at state $s_t$ if and only if there exists an action $a^i\in A^i$ such that:
  \begin{equation}
  \label{eq:IR}
 \mathbb{E}_{\mathcal{M}_\text{pa}}^{\boldsymbol{\pi}}\left[\left(b_t(\theta^i r^i_t) - c \right) + V^i_{t+1}(s_{t+1}) ~\middle|~ s_t, b_t, a_t^i = a^i \right] \geq 0.
  \end{equation}
\end{definition}

\subsection{Linear State-Dependent Contracts}
\label{subsec:fair_contracts}

\begin{algorithm}[t]
\caption{Principal-Agent Policy Gradients with Linear Contracts}
\label{alg:contract_pg}

\begin{algorithmic}[1]
\STATE \textbf{Input:} $(n+1)$-player principal-agent Markov game ${\mathcal{M}_\text{pa}}$; learning rates $\eta_\text{p}$ (principal), $\eta_\text{a}$ (agents)
\STATE Randomly initialize policy parameters $\phi_0^1, \dots, \phi_0^{n}, \phi_0^{p}$
\FOR{$k = 0, \dots$}
  \FOR{episode $e = 1, \dots, N$}
    \STATE Reset game ${\mathcal{M}_\text{pa}}$ to initial state $s_0$
    \STATE Initialize episodic wealth $\hat{w}_e^i \gets 0$ for all players
    \FOR{timestep $t = 1, \dots, T$}
      \STATE Principal selects contract $\alpha_t \sim \pi_\text{p}(\cdot \mid s_t; \phi_k^p)$
      \STATE Each agent $i\neq p$ selects action $a_t^i \sim \pi^i_\text{a}(\cdot \mid s_t, \alpha_t; \phi_k^i)$
      \STATE Observe rewards $\mathbf{r}_t$ and next state $s_{t+1}$
      \FOR{agent $i= 1, \dots, n$}
        \IF{$a^i_t \neq \text{reject}$}
          \STATE Pay agent $\hat{w}_e^i \gets \hat{w}_e^i + \left(\alpha_t \theta^i r_t^i - c\right)$
          \STATE Collect $\hat{w}_e^p \gets \hat{w}_e^p + (1-\alpha_t) \theta^i r_t^i $
        \ENDIF
      \ENDFOR
    \ENDFOR
  \ENDFOR
  \STATE Estimate wealth $\hat{w}^i \gets \frac{1}{N} \sum_{e=1}^N \hat{w}_e^i$ for each player $i$
  \STATE {\color{gray} // Gradients can be approximated through any policy-gradient methods (e.g. PPO)}
  \STATE Update principal's policy: $\phi_{k+1}^p \gets \phi_k^p + \eta_\text{p} \nabla_{\phi_p} \hat{w}^p$
  \STATE Update each agent's policy : $\phi_{k+1}^i \gets \phi_k^i + \eta_\text{a} \nabla_{\phi_i} \hat{w}^i, \;i \neq p$
\ENDFOR
\end{algorithmic}
\end{algorithm}

\emph{Linear contracts} form a particularly interpretable and practical class of contracts, in which agents receive a proportion of the rewards they help generate, rather than a fixed payment. In this setup, a contract is parameterized by a single scalar $\alpha_t \in [0, 1]$, representing the agent's share of the reward. Linear contracts are not only easier to analyze and interpret, but they also admit efficient geometric solutions for optimal design~\cite{duetting_simple_2019}. Additionally, it ensures that LL always holds, as the principal always pays agents and not the opposite ($\alpha \geq 0$). 
For these reasons, we restrict the principal to issuing linear contracts, $b_t(\theta^i r_t^i) = \alpha_t \theta^i r_t^i$. Let $S_t = \sum_{i=1}^{n} \theta^i r^i_t\cdot \mathds{I}[a^i_t \neq \varnothing]$ be the sum of the rewards collected by the agents. Under this formulation, the objectives of the principal and agents become:
\begin{align}
  \label{eq:objectives}
  \max_{\pi_\text{p}} \mathbb{E}^{\boldsymbol{\pi}}_{\mathcal{M}_\text{pa}} & \left[ \sum_{t=0}^{T-1} (1 - \alpha_t) S_t  \;\middle|\; s_0 \right], 
  \\
  \max_{\pi^i_\text{a}} \mathbb{E}^{\boldsymbol{\pi}}_{\mathcal{M}_\text{pa}} & \left[ \sum_{t=0}^{T-1} \left(\alpha_t \theta^i r^i_t - c \right) \cdot \mathds{I}[a^i_t \neq \varnothing] \;\middle|\; s_0 \right].
\end{align}

\begin{lemma}[Budget balance of the linear contracts]
    \label{le:balance}
    Let $n_k \triangleq \sum_i \mathds{I}[a_k^i \neq \varnothing]$ be the number of agents accepting the contract at timestep $k$. At each timestep $t$, for any sequence of contracts $(\alpha_0, ..., \alpha_t) \in [0,1]$, uniform costs $c$ and any outputs realized by agents, the welfare $W_t$:
    \begin{align}
        W_t &= w^p_t + \sum_i w^i_t = \sum_{k=0}^t \left[ R_p(\cdot, \alpha_k) + \sum_i R_a^i(\cdot, \alpha_k)\right]\notag\\
            &= \sum_{k=0}^t \left[ (1 - \alpha_k) S_k + \sum_i \left( \alpha_k \theta^i r_k^i - c \right) \cdot \mathds{I}[a^i_k \neq \varnothing] \right] \notag\\ 
            &= \sum_{k=0}^t \left[ S_k - \alpha_k S_k + \alpha_k S_k - cn_k \right]\notag\\ 
            &= \sum_{k=0}^t \left[ S_k - cn_k \right].
    \end{align} 
    is independent of $\alpha$. 
\end{lemma}

\begin{lemma}
    \label{le:affine}
    Rewriting reward as an affine function of contract for each player $i$:
    \begin{align*}
        R^i_t = x_t^i \alpha_t + y_t^i
    \end{align*}
    The sum of linear coefficients $x_t$ is equal to 0.
\end{lemma}

Lemma~\ref{le:balance} forms an analogue to the concept of budget balance in the field of mechanism design~\cite{green_laffont_1979} - the linear contract acts as a redistributing mechanism, and it indeed preserves welfare for the fixed actions. The contract affects the welfare indirectly, through the acceptance and the effort induced.

\subsection{Learning Contracts with Policy Gradients}
\label{sec:policy_gradients}

We address this learning problem using \emph{policy gradient} methods, treating both the principal and agents as learners. In particular, we model the principal's policy as a Gaussian distribution: contracts at state $s$ are sampled from a normal distribution $\mathcal{N}(\mu_s, \sigma_s)$, where $\mu_s$ and $\sigma_s$ are state-dependent parameters learned by the principal. This makes the policy differentiable and allows for exploration.

Algorithm~\ref{alg:contract_pg} outlines the learning procedure in this setting. After collecting sufficient experience to estimate policy gradients, the principal and agents update their policies. This interaction can be viewed as a bilevel optimization, since principal-agent interactions form a sequential Stackelberg game~\cite{ivanov_principal-agent_2024}, where agents are expected to best-respond to the contracts offered. If the roles were reversed and instead the principal best responded to the agents’ current strategies, she would immediately propose degenerate contracts with $\alpha=0$, before the agents could learn to reject contracts that violate the IR constraint, leaving them with no incentive to accept the offer and collapsing the system. Avoiding this requires the principal to learn on a slower timescale, following the results from two-timescale stochastic approximation~\cite{borkar_1997} and the actor-critic methods~\cite{konda_1999}. We accordingly set $\eta_\text{p} \ll \eta_\text{a}$.

Even under balanced learning dynamics, a principal that seeks only to maximize her own wealth may converge to offering contracts that barely satisfy the IR constraint, leaving agents with near-zero returns. While such contracts seem optimal from a purely greedy perspective, they are problematic for two key reasons: (1) they are inherently unfair, and (2) they diminish agents’ incentives to explore and improve, ultimately reducing system performance. In practice, a principal therefore benefits from designing fairer contracts, not only to maintain cooperation but also to encourage agent engagement and adaptation. Ultimately, we want to improve the system's fairness by introducing a third party coordinating the work of the multiple independent agents: a role a purely greedy principal cannot fulfill. This motivates the need to regularize the principal’s objective, either by directly promoting agents' welfare or by penalizing large disparities in wealth between all parties.

\section{Regularization for Fair Contracts}
\label{sec:regularization}

Rather than solely maximizing her own wealth, the principal can
incorporate fairness considerations into her objective. Promoting fairness requires some degree of altruism from the
principal, which can be quantified by an altruism
parameter $\lambda$~\cite{sawyer:altruism:1966}. This captures the
extent to which she values the overall fairness of the
system relative to the profits.

\begin{definition}[Fairness regularization]
    Let $F$ be a specific fairness function. Then the regularized objective of the principal takes form:
    \begin{align}
        J(s_t, \mathbf{a}_t, s_{t+1}, b_t) = \sum_{i=1}^n (1 - \alpha_t) \theta^i r_t^i + \lambda F
    \end{align}
\end{definition}

The coefficient $\lambda$ plays a central role by modulating the degree of altruism in the principal's behavior. When $\lambda \rightarrow 0$, the principal greedily maximizes her own wealth, leaning towards the contract policy which barely allows agents to offset their costs. As $\lambda$ grows, the principal steadily becomes altruistic, up to a point where she becomes a central planner, disregarding her own wealth entirely. 

We examine two possible fairness regularizers. First, in Section~\ref{subsec:welfare_reg}, as a baseline, we adapt a welfare-based regularization approach, inspired by the formulation of~\cite{ivanov_principal-agent_2024}. Then, in Section~\ref{subsec:fairness_reg}, we present an alternative regularization method that penalizes disparities in the wealth accumulated by all parties, using variance as a fairness metric. Not only do the regularizations introduced in this section modify the reward function, but as a result, the principal's learning objective changes.

\subsection{Welfare-based regularization}
\label{subsec:welfare_reg}

It has been shown that directly tying the principal's objective to agent wealth can encourage more cooperative behavior~\cite{ivanov_principal-agent_2024}.  Following this insight, we incorporate system welfare into the principal's reward by augmenting it with the agents' collected rewards. After simplifying, this leads to a modified reward function:
\begin{align}
  \label{eq:welfare_regularization}
J^{\text{welfare}}(s_t, \mathbf{a}_t, s_{t+1}, b_t) = \sum_{i=1}^{n}  (1-\alpha_t + \lambda)\theta^i r_t^i.
\end{align}
Here, the principal observes the rewards $\theta^i r_t^i$ for each agent but cannot infer the agent's underlying type. This regularization does not require keeping a history of rewards and is intentionally simple. However, it does not account for the heterogeneity of the environment, therefore the system's final outcomes might not be equitable.

\subsection{Variance-based regularization}
\label{subsec:fairness_reg}

An alternative approach is to regularize the principal's objective based on the distribution of wealth, rather than overall welfare. Specifically, we suggest penalizing the principal for creating inequalities in the accumulated wealth across all parties, thereby encouraging contract offers that promote a more balanced distribution over an episode. The simplest form of fairness endorsing equity would be to just use negative variance in wealth: $F(\mathcal{W}) = -\Var[\mathcal{W}]$. This formulation is simple and computationally tractable. On top of that, the variance regularizer allows us to compute a \textit{post factum} optimal contract for each timestep, serving as a benchmark.

\begin{remark}
    \label{re:cumwealth}
    Cumulative wealth at time $t$ is given by:
    \begin{align*}
        w^i_t(\alpha_t) &= (w^i_{t-1}-c \cdot \mathds{I}[a^i_t \neq \varnothing]) + \theta^ir_t^i \alpha_t, \\ \notag w^p_t(\alpha_t) &= (w^p_{t-1}+S_t) - S_t\alpha_t.
    \end{align*}
\end{remark}

\begin{lemma}[Variance minimizer]
    \label{le:varmin}
    Let $A_t \triangleq \frac{1}{n+1} \left( S_t^2 + \sum_i (\theta^i r_t^i)^2 \right)$ and 
    \begin{align*}
        B_t \triangleq \frac{2}{n+1} \left[ -S_t (w_{t-1}^p + S_t) + \sum_i \left( \theta^i r_t^i (w_{t-1}^i - c \cdot \mathds{I}[a^i_t \neq \varnothing]) \right) \right]
    \end{align*}
    Variance of the wealth of all players is then represented by:
    \begin{align}
        \Var(\mathcal{W}_t(\alpha_t)) = A_t \alpha_t^2 + B_t \alpha_t + C_t
    \end{align}
    for some constant $C_t$. Moreover, the contract minimizing the variance at a given step is given by:
    \begin{align}
        \label{eq:variance_minimizer}
        \alpha^{Vmin}_t \triangleq - \frac{B_t}{2 A_t}
    \end{align}
    constrained by $\alpha \in [0,1]$.
\end{lemma}

\textit{Proof sketch. } Following Lemma~\ref{le:affine}, the variance can be expanded as 
\begin{align*}
    \Var(\alpha_t) &= \frac{1}{n+1} \sum_{k=1}^{n+1} \left( w_t^k(\alpha_t) - \bar{w}_t(\alpha_t) \right)^2\\ &= \frac{1}{n+1} \sum_{k=1}^{n+1} \Big[ \alpha_t^2 (x_k - \bar{x})^2 + 2 \alpha_t (x_k - \bar{x}) (y_k - \bar{y}) \\ &\qquad\qquad\qquad + (y_k - \bar{y})^2 \Big]
\end{align*}

Since $\sum_k x_k = 0$, it implies that its mean $\bar{x} = 0$. This, after simplifying, yields the $A_t$ and $B_t$ terms from Lemma~\ref{le:varmin}. Applying simple algebra, we end up with the minimizing linear contract from Equation~\ref{eq:variance_minimizer}.\hfill$\square$

\begin{theorem}[One-step optimal contract for fairness regularization]
  \label{thm:maximizer}
  The principal's objective $J$ at timestep $t$, regularized with $F(\mathcal{W}) = - \Var(\mathcal{W})$ is maximized by:
  \begin{align}
    \alpha_t^*(\lambda) = \alpha_t^{Vmin} - \frac{S_t}{2 A_t \lambda}
  \end{align}
  which approaches $\alpha^{Vmin}_t$ as $\lambda \rightarrow \infty$.
\end{theorem}

\textit{Proof.} Rewriting $J = S_t - S_t \alpha_t - \lambda(A_t \alpha_t^2 + B_t \alpha_t + C_t)$. The function is quadratic and concave, thus it has a unique maximizer, obtained by setting its first order derivative $J' = -S_t - 2 A_t \lambda \alpha_t - B_t \lambda = 0$. \hfill $\square$

Theorem~\ref{thm:maximizer} highlights the role of the altruism parameter $\lambda$ - as $\lambda \rightarrow 0$, the optimal contract approaches the greedy one, and the variance minimizing one when $\lambda \rightarrow \infty$. The term $\frac{S_t}{2A_t\lambda}$ quantifies how much the principal is lowering the contract offer for her own profit.

In the learning setting, however, the value of $\alpha^*_t(\lambda)$ serves as a benchmark rather than an implementable policy. All the results above rely on the premise that the principal knows upfront the reward produced by the agents at the current state. This premise concerns the analytical benchmark only, while in Algorithm~\ref{alg:contract_pg} the contract $\alpha_t$ is chosen before the rewards are realized, and the rewards enter the principal's objective only as a learning signal. In the case when the principal learns the state-dependent policy, the optimal contract is realizable only to the extent that $s_t$ is informative about the past wealth distribution among all parties. Additionally, Theorem~\ref{thm:maximizer} treats agents' actions as fixed and returns an optimal value for the single timestep, while in the learning setting agents respond to the offered contract in a Stackelberg fashion, with the principal optimizing the objective over the course of the episode. A discrepancy between the learned policy and myopic one-step optimum is therefore expected.

Finally, one can ask why is there a need to incorporate the principal's gain in the objective at all. After all, by Theorem~\ref{thm:maximizer}, if $\lambda \rightarrow \infty$, we should approach the variance-minimizing contract from Equation~\ref{eq:variance_minimizer}. 

\begin{proposition}
  \label{prop:nontrivial}
  Let $\mathcal{D}$ be the set of equal wealth vectors - $\mathcal{D} \triangleq \left\{ w \in \mathbb{R}^{n+1} : w^1 = ... = w^{n} = w^p \right\}$, so $\forall_{\mathcal{W} \in \mathcal{D}} \Var(\mathcal{W}) = 0$. If $J = -\Var(\mathcal{W})$, the principal prefers to offer contracts not meeting the IR criteria of the agents, that result in at least as good utility for the principal as the contracts actually inducing production.
\end{proposition}

Taking this into account, the immediate reward of the principal in the objective $(1-\alpha) S_t$ indirectly incentivizes production, and in consequence, the welfare of the entire system. This helps stabilize the learning process, which otherwise would have an incentive to collapse into one of the suboptimal solutions from $\mathcal{D}$. 

\begin{table*}[t]
  \caption{Resulting scores of the Coin Game environment. *For NoP, metrics were computed without the principal.}
  \label{tab:training} 
  \centering
  \begin{tabular*}{\textwidth}{@{\extracolsep{\fill}}lcccccccccc} 
    \toprule
    &NoP* & Greedy & Fix & \multicolumn{7}{c}{Regularized} \\
    \cmidrule{5-11}
    && & & \multicolumn{4}{c}{Welfare} & \multicolumn{3}{c}{Wealth Variance}\\
    \cmidrule(r){5-8}\cmidrule(l){9-11}
    $\lambda$ & & & &  1 & 9 & 12 & 15 & 0.75 & 1 & 1.25 \\
    \midrule
    Variance & 116.7 & \textbf{4.7} & 17.9 & 77.9 & 81.4 & 73.0 & 6.9 & 19.2 & 13.7 & 14.0\\
    \midrule
    Welfare & 44.6  & 3.4  & \textbf{45.5}  &  28.6  & 34.4  & 43.0 & 21.2 & 44.6 &  \textbf{45.5}  & 44.7 \\
    \midrule
    1 - Gini & 0.952  & 0.200 & 0.947 & 0.704 & 0.834 & 0.852 & 0.957 & 0.978 & \textbf{0.992} & 0.986 \\ 
    \midrule
    Rawlsian & 18.0  & -0.2  & 11.2 &  0.9  & 5.9  & 6.9 & 5.7 & 13.3 & \textbf{14.7} & 14.1 \\
    \midrule
    AIE & 42.5  & 0.7  & 43.1  & 20.3  & 29.9  & 36.8 & 20.3 & 43.6 & \textbf{45.1} & 44.1 \\

    \bottomrule
  \end{tabular*}
\end{table*}
\begin{figure*}[t]
    \centering
    
    \begin{subfigure}[b]{0.45\textwidth}
        \includegraphics[width=\textwidth]{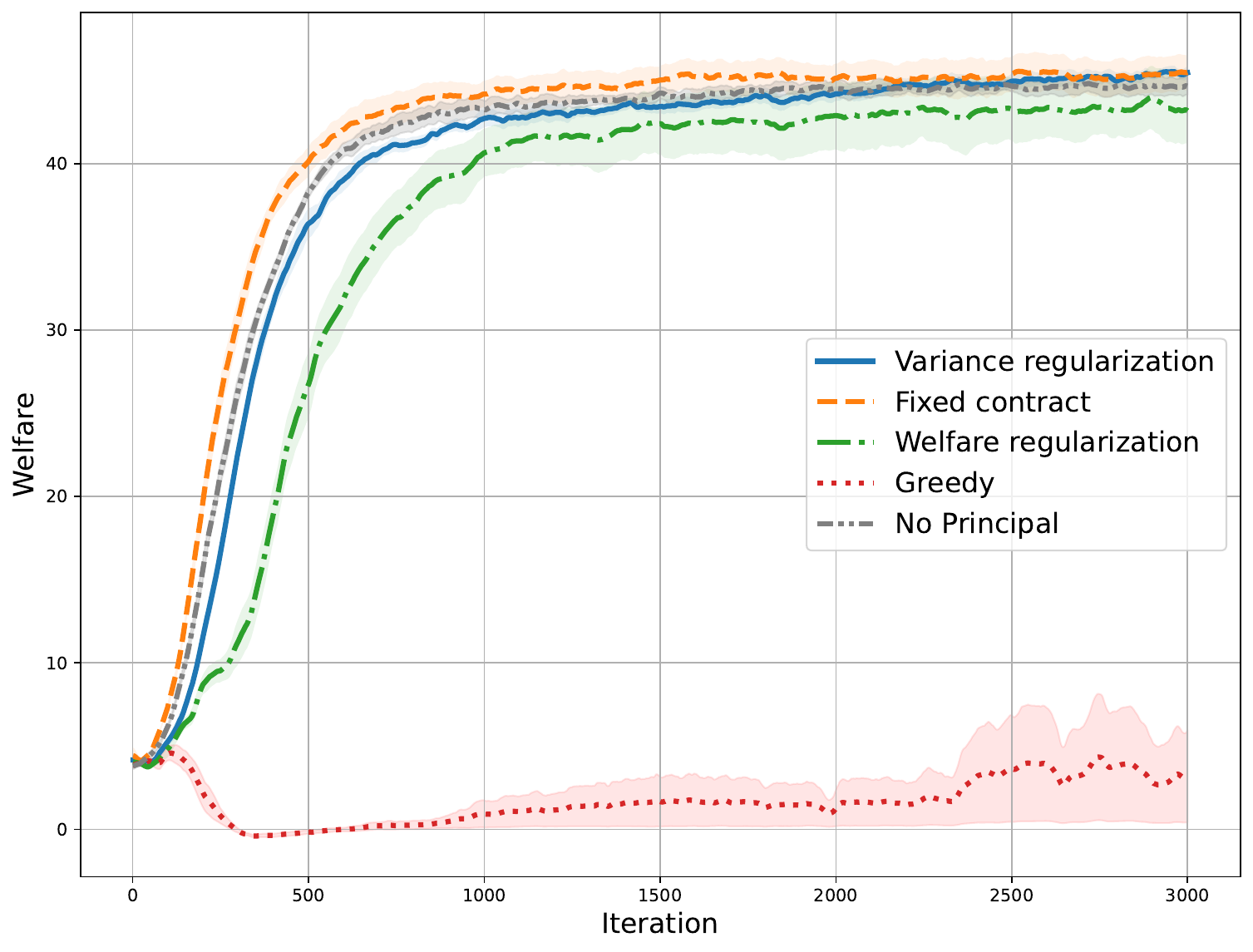}
        \caption{}
        \label{fig:welfare}
    \end{subfigure}
    \hfill 
    \begin{subfigure}[b]{0.45\textwidth}
        \includegraphics[width=\textwidth]{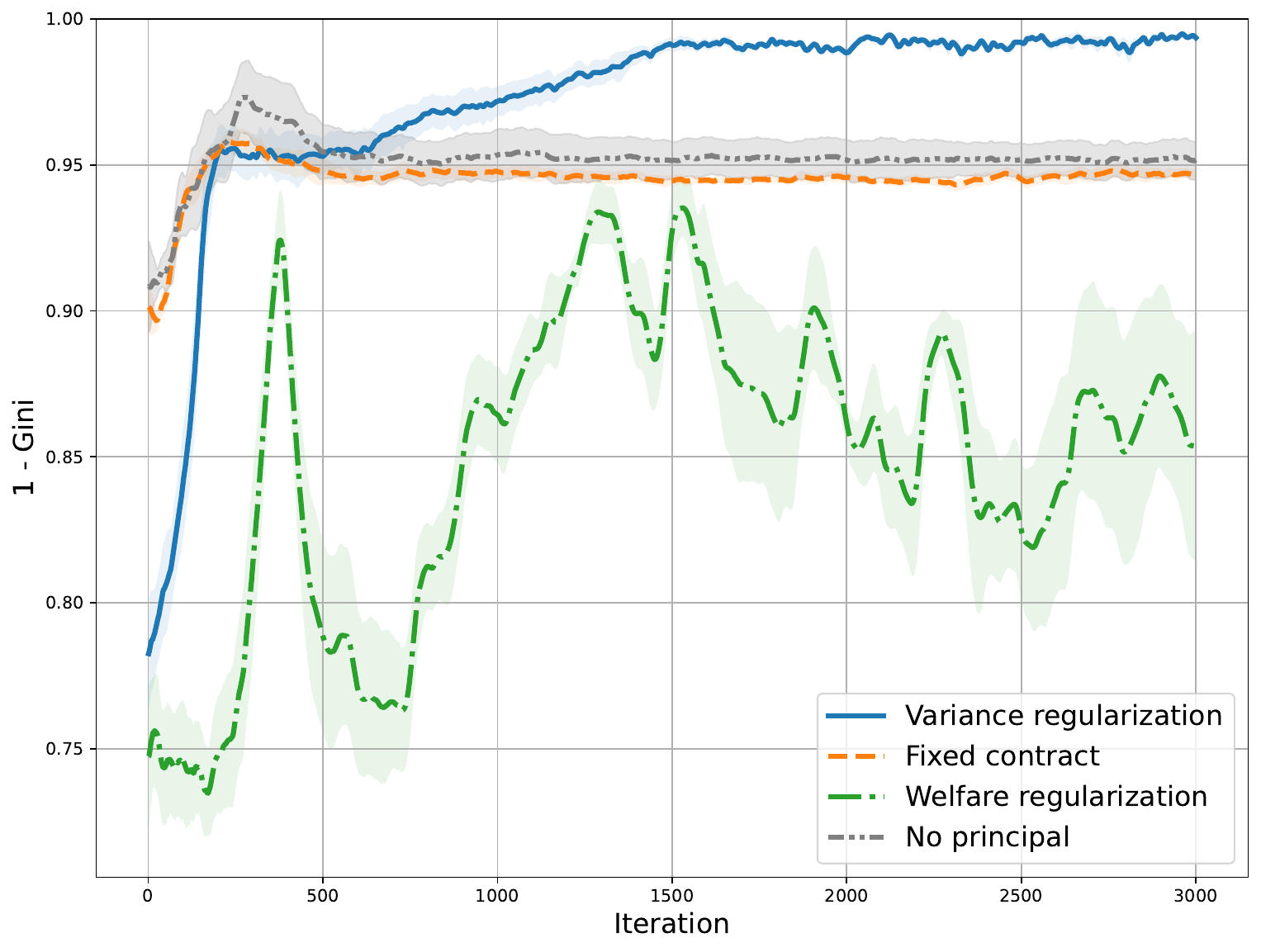}
        \caption{}
        \label{fig:gini}
    \end{subfigure}
    \caption{Results of the Coin Game experiments. (a) Comparison of average social welfare attained during the learning process by each benchmark. (b) Comparison of equity measured by 1 - Gini metric across benchmarks. The Greedy benchmark was omitted from this plot for the sake of readability, since it achieved noticeably lower score than the rest. In both plots WR had $\lambda = 12$ and VR used $\lambda = 1$.
}
    \label{fig:grid_coin}
    \Description{Two line plots, comparing the welfare and 1-Gini metric across all benchmarks in the Coin Game.}
\end{figure*}

\section{Experiments}
\label{sec:experiments}

The main goal of our experiments is to see if fairness-aware principals can promote agent welfare without sacrificing much of their utility. We examine whether a linear homogeneous contract, conditioned only on the current observation, can arrive at a stable state with equal wealth allocation. To do so, we examine how effective regularization is, relative to three baselines, namely:
\begin{enumerate}
  \item No Principal (NoP): equivalent to a principal with fixed linear contracts $\alpha=1$. In the case of this baseline, all the statistics are computed for agents only, to provide insight into their behavior without the principal. As a result, only welfare is directly comparable with other benchmarks.
  \item Greedy Principal (Greedy): the principal has no form of regularization, therefore aiming to maximize her own returns only.
  \item Fixed contracts (Fix): With the number of agents $n=2$, $ \alpha = \frac{n}{n+1} = \frac{2}{3}$ is the fixed contract value given at each timestep, equalizing the wealth of the parties provided that the agents are homogeneous. Here, the principal always retains one-third of the rewards, distributing the remainder between the agents. This benchmark evaluates the performance of a heuristic approach designed for a homogeneous population when applied to a heterogeneous one. 

\end{enumerate}

\begin{table*}[t]
  \caption{Resulting scores of the Gather-Trade-Build environment. *For NoP, metrics were computed without the principal.}
  \label{tab:training_aie} 
  \centering
  \begin{tabular*}{\textwidth}{@{\extracolsep{\fill}}lccccccccc} 
    \toprule
    &NoP* & Greedy & Fix & \multicolumn{6}{c}{Regularized} \\
    \cmidrule{5-10}
    && & & \multicolumn{3}{c}{Welfare} & \multicolumn{3}{c}{Wealth Variance}\\
    \cmidrule(r){5-7}\cmidrule(l){8-10}
    $\lambda$ & & & &  1 & 10 & 100 & 0.1 & 1.0 & 1.5  \\
    \midrule 
    Variance & 1481.0 & 736.2 & 946.2 & \textbf{442.5} & 1380.5 & 850.4 & 2161.5 & 515.6 & 829.7\\ 
    \midrule
    Welfare & 171.0  & 134.5  & 180.7  &  172.3  & 189.9  & 146.3 & \textbf{306.4} & 152.3 & 173.7  \\
    \midrule
    1 - Gini & 0.929  & 0.854 & 0.917 & 0.882 & 0.864 & 0.782 & 0.913 & \textbf{0.922} & 0.875 \\ 
    \midrule
    Rawlsian & 63.0  & 17.7  & 38.3 &  35.2  & 31.5  & 6.3 & \textbf{75.5} & 33.1 & 32.4 \\
    \midrule
    AIE & 159.8  & 116.4  & 166.1  & 152.0  & 168.2  & 114.4 & \textbf{288.4} & 140.6 & 152.1 \\
    \bottomrule
  \end{tabular*}
\end{table*}

\begin{figure*}[t]
    \centering
    
    \begin{subfigure}[b]{0.45\textwidth}
        \includegraphics[width=\textwidth]{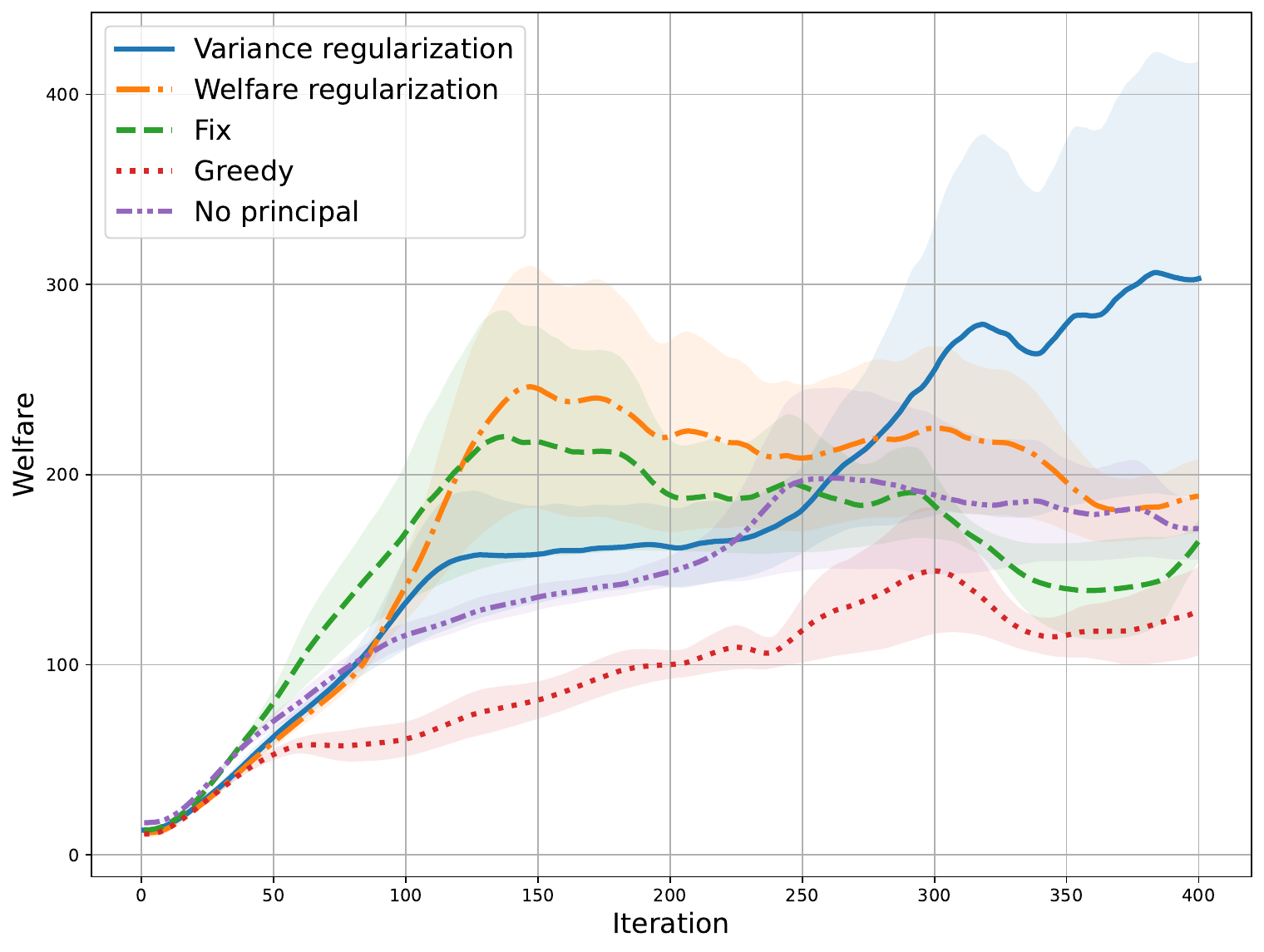}
        \caption{}
        \label{fig:welfare_aie}
    \end{subfigure}
    \hfill 
    \begin{subfigure}[b]{0.45\textwidth}
        \includegraphics[width=\textwidth]{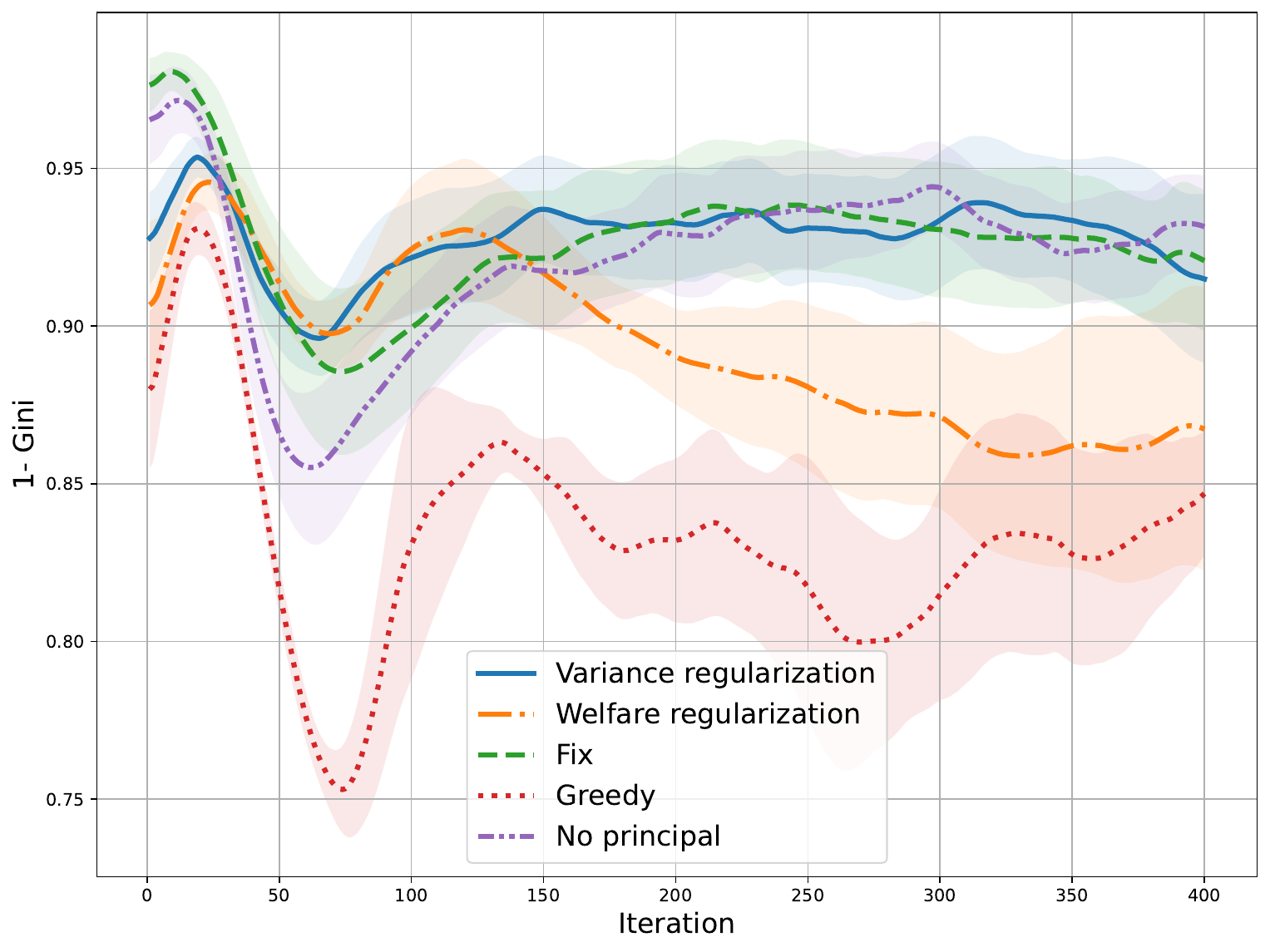}
        \caption{}
        \label{fig:gini_aie}
    \end{subfigure}

    \caption{Results of the Gather-Trade-Build experiments.  (a) Comparison of average social welfare attained during the learning process by each benchmark. (b) Comparison of equity measured by 1 - Gini metric across benchmarks. In both plots WR had $\lambda = 10$ and VR used $\lambda = 0.1$ in their regularization terms.}
    \label{fig:grid_aie}
    \Description{Two line plots, comparing the welfare and 1-Gini metric across all benchmarks in the Gather-Trade-Build.}
\end{figure*}

While agents seek to maximize their individual welfare under the contract dynamics, the principal's goal, when regularized, is to design contracts that promote welfare for agents (WR) or equitable welfare distribution (VR). Given that multiple metrics for measuring such equity have been proposed in the literature, we include several of them in our study:
\begin{enumerate}
  \item Variance between the wealth of the principal and the agents, as used in the regularization of the VR benchmark.
  \item Welfare: unlike variance, it provides us with the information whether fairness comes at the price of welfare when compared to other solutions.
  \item The $1 - \text{Gini}$ index: based on the widely used parity metric of Gini index~\cite{gini1912}. It maps equality in wealth to the range $[0,1]$, from the most unequal to most equal:
\begin{gather*}
  1 - \text{Gini} = 1 - \frac{1}{2(n+1)W_T} \sum_i^{n+1} \sum_j^{n+1} |w_i - w_j|,
\end{gather*}
  while capturing similar properties as variance, it is unit-independent.
  \item The Rawlsian index~\cite{rawls1971,zheng_ai_2020}: the wealth of the poorest player : $\min_{w_T \in \mathcal{W}_T} w_T$.
  \item The product of the $1 - \text{Gini}$ index and welfare: a metric used in the AI Economist~\cite{zheng_ai_2020}, which we call AIE. It corresponds to the product of equality and productivity, capturing both desired properties. 
\end{enumerate}
All metrics are computed over all parties, including the principal. After all, measuring fairness between the agents only would deem a principal retaining all the wealth fair. In all of our experiments and aligning with Section~\ref{subsec:fair_contracts}, principals always learn linear contracts with shares within $[0,1]$. Learning for both the principal and the agents is done via Proximal Policy Optimization (PPO). All metrics are averaged over 5 random seeds.

\subsection{Coin Game} 
\label{subsec:coin_game}

We consider a modified form of the Coin Game~\cite{foerster_learning_2018}, a sequential social dilemma that considers two agents, red and blue, who move around a square grid. Their goal is to collect a coin, which can also be red or blue. The coin is always present on one of the unoccupied grid spots. Upon collection, a new coin is generated randomly on the grid. The agents are rewarded with one point if they collect a coin matching their color. If they collect a coin not matching their color, they are still rewarded with 0.2 points. The reward is then scaled by their type.
Selfish agents will pay no attention to the wealth of another agent, seeking to maximize their own. It is detrimental to the welfare of the entire system, with less skilled agents having fewer incentives to move and stagnate. To impose fair dynamics between the red and blue agents, any ties are broken at random, and agents can occupy the same square in the grid.

\paragraph{Results.}

The results are presented in Table~\ref{tab:training} and Figure~\ref{fig:grid_coin}. VR outperforms all other benchmarks, consistently achieving near-optimal $1 - \text{Gini}$ score, while achieving very high welfare at the same time, which in turn is highlighted by the highest AIE metric. Moreover, the Rawlsian metric is the highest amongst the benchmarks with three players (NoP consists of two agents only). While WR achieves good results in terms of welfare, it is quite unstable, not presenting the convergent behavior on equity that Fix, NoP, and VR have. Lastly, we observe that greedily learning contracts leads to a system collapse in this environment.

\subsection{Gather-Trade-Build}

To evaluate how our approaches perform in more complex environments, we conducted further experiments on a modified version of the Gather-Trade-Build environment introduced in the AI Economist~\cite{zheng_ai_2020}.
In this setting, multiple independent agents interact by collecting resources. These resources can either be traded for money with other agents or used to build houses in exchange for rewards. Importantly, the reward depends on the agent’s \emph{skill}, which remains private information. This notion of skill satisfies the definition of agent type introduced in our paper. The environment is also partially observable for both principal and agents. Inventories (i.e., the amount of money and resources held by each agent) remain hidden from the principal and other agents. This limited information makes coordination between the principal and agents more challenging. To make the results comparable with the Coin Game experiments, we have set the number of agents to 2.

\paragraph{Results. } Table~\ref{tab:training_aie} and Figure~\ref{fig:grid_aie} summarize the results in the Gather-Trade-Build environment. The VR method achieves the highest welfare, AIE, and Rawlsian scores across all benchmarks, although it exhibits high variability across the seeds during the later stages of training (Figure~\ref{fig:welfare_aie}).
Notably, its strong AIE performance indicates that while NoP and Fix obtain comparable results in terms of 1-Gini, VR is the only approach that simultaneously promotes both Rawlsian and AIE. In contrast, the greedy approach again produces the least desirable outcomes.

\paragraph{Distance to the optimal contract.} In both environments, the learned policies result in high discrepancies between the contract value offered by the principal and the one-step optimum resulting from Theorem~\ref{thm:maximizer}. This is expected, since the one step optimum is computed using the information unavailable at decision time, while Algorithm~\ref{alg:contract_pg} optimizes the episodic objective, ascending on the cumulative wealth distribution.

\section{Conclusion and Future Work}
\label{sec:conclusion}

We have proposed a framework for designing fair contracts in SSD games with heterogeneous agents and verified the results empirically. Crucially, we have shown that simple linear fairness-aware contracts are able to induce cooperative behavior between agents, improving the fairness of the entire system, without sacrificing welfare in comparison to the baselines.

\paragraph{Limitations.} Our experiments are limited to simple domains, which may not capture the complexities of real-world settings. Additionally, while linear contracts offer simplicity and interpretability, it is not necessarily the case for learned contract policies, which remain difficult to analyze.

\paragraph{Future Work.} A natural extension of this work involves exploring more expressive contract forms, where payments can be conditioned on richer outcome spaces. While potentially harder to interpret, such contracts may offer greater flexibility and effectiveness in steering the system toward desired outcomes. Instead of a grid search, the altruism parameter $\lambda$ might be set in an automatic or an adaptive manner. Another promising direction is to consider deceptive agents who might try to game the contracts offered by the principal, and how such behavior could be counteracted. Finally, it would be interesting to explore how adaptive contracts could be leveraged when the principal's goal is to allocate wealth according to an arbitrary target distribution rather than ensuring equity.

\begin{acks}
This research was partially supported by the Swiss National Science Foundation (grant $\text{\textnumero}\,$ 219499).
\end{acks}

\bibliographystyle{ACM-Reference-Format}
\bibliography{FairCT}

\end{document}